\documentclass[twocolumn,aps,prd,groupedaddress,showpacs]
              {revtex4}%
\usepackage{graphicx}
\usepackage{amsmath}
\usepackage{bm}
\usepackage{relsize}
\RequirePackage{xspace}
\newcommand{\bqa}{\begin{eqnarray}}
\newcommand{\eqa}{\end{eqnarray}}
\newcommand{\pslash}{\slash\hspace{-0.55em}}

\begin{document}


\title{\mbox{}\\[10pt]
$\Upsilon$ radiative decays to light quark jets and color octet
mechanism}

\author{Ying-Jia Gao$~^{(a)}$,Yu-Jie Zhang$~^{(a)}$, and Kuang-Ta Chao$~^{(b,a)}$}
\affiliation{ {\footnotesize (a)~Department of Physics, Peking
University,
 Beijing 100871, People's Republic of China}\\
{\footnotesize (b)~China Center of Advanced Science and Technology
(World Laboratory), Beijing 100080, People's Republic of China}}




\begin{abstract}
We study radiative decays of $\Upsilon$ to light quark jets in
nonrelativistic QCD by taking both the color singlet and color
octet $b\bar b$ operators into consideration. The cut for quark
jet energy and cut for the angle between two quark jets are
introduced. The sensitivity to the soft and collinear
singularities in the loop integrals are greatly reduced by these
cuts. With the jet energy cut of about 1~GeV, and the jet angle
cut of about $36^\circ$, the branching ratio for
$\Upsilon\to\gamma q\bar q$ is found to be $8.2\times 10^{-4}$
from color singlet contributions. The color octet contributions
could be much larger than that of color singlet, depending on the
estimate of the color octet matrix elements. This process may
provide a new test for the color octet mechanism in
nonrelativistic QCD.
\end{abstract}


\maketitle


\section{INTRODUCTION}

The observation of $\Upsilon$ decay into three-gluon jets
$\Upsilon\to ggg$ has been viewed as a crucial event for verifying
quantum chromodynamics (QCD). The radiative decay $\Upsilon\to
\gamma gg$ has also been an important testing ground for QCD.
These processes are computable in perturbative QCD due to the
large mass of the $b$ quark. Recently the CLEO Collaboration has
accumulated a large number data sample for the low lying
$\Upsilon$ resonances \cite{cleo1, cleo2}. This encourages us to
study the radiative decay process $\Upsilon\to\gamma q\bar q$,
where $q$ and $\bar q$~$(q=u,d,s)$ are referred to as light quark
jets.

Jets are considered to be footprints of quarks and gluons produced
at short distances. While the gluon jet, its production angular
distribution, and the determination of the gluon spin have
recently been studied\cite{gluonex}, less attention has been paid
to quark jets, say, e.g. in $\Upsilon$ decays. Since the gluon jet
is expected to be "fatter" than the quark jet\cite{alam} or the
mean multiplicity of the gluon jet is expected to be larger than
the quark jet (see e.g. \cite{brod}), it will be interesting to
measure the quark jets production in comparison with the gluon
jets in $\Upsilon$ radiative decays.

In heavy quarkonium production and decay an important issue is about
the color octet mechanism in nonrelativistic QCD (NRQCD)\cite{BBL},
which is introduced to explain the large cross sections of inclusive
charmonium hadroproduction at large $p_T$ in $p\bar{p}$ collisions
measured at the Fermilab Tevatron\cite{cdf}, and other decay and
production processes. Though the color octet mechanism has gained a
series of successes, some difficulties still remain and more tests
are needed (for a recent comprehensive review, see
\cite{Brambilla:2004wf}). In this regard, the radiative decay
process $\Upsilon\to\gamma q\bar q$ may provide an interesting test
for the color-octet mechanism, since the color-octet $b\bar b$
component (with soft gluons) in the $\Upsilon$ wave function could
give substantial contributions to this process (for discussions on
color-octet contributions in $\Upsilon$ hadronic and radiative
decays, see, e.g. \cite{CKY,fleming1,co4}, see
also\cite{Brambilla:2004wf} for more references).

Based on the above considerations, we will study the three body
decay $\Upsilon \to \gamma q \bar{q}$, where $q=u,d,s$. We use the
$b\bar b$ bound state description for $\Upsilon$ in NRQCD and give
the results through both color-singlet and color-octet $b\bar b$
operators. In NRQCD, with the Fock state expansion in terms of
$v$, the relative velocity between $b$ and $\bar b$, the decay
width for $\Upsilon(1S)$ can be written as a sum of contributions
from various $b\bar b$ channels with different color and angular
momentum:
 \bqa
 d\Gamma[\Upsilon(1S)]&=&d\Gamma[b\bar{b}_1(^3S_1)]\langle
 \Upsilon|\mathcal{O}_1(^3S_1)|\Upsilon\rangle
 +d\Gamma[b\bar{b}_8(^3S_1)]\nonumber\\
 &\times&\langle \Upsilon|\mathcal{O}_8(^3S_1)|\Upsilon\rangle
  +d\Gamma[b\bar{b}_8(^1S_0)]\nonumber\\
  &\times&\langle \Upsilon|\mathcal{O}_8(^1S_0)|\Upsilon\rangle
 +\bigl(\sum_J(2 J+1)\nonumber\\
 &\times&d\Gamma[b\bar{b}_8(^3P_J)]\bigr)\langle
 \Upsilon|\mathcal{O}_8(^3P_0)|\Upsilon\rangle,
 \eqa
where the long distance matrix elements with different color
(denoted by the subscript $1(8)$ for color-singlet (octet)) are
involved. In Section 2 we will describe our method and give a
calculation for the color singlet contribution. In section 3, we
will study color octet contributions. Then we will give a summary
in Section 4.

\section{ Color singlet contributions}

 \begin{figure*}[t]
\includegraphics[height=7cm]{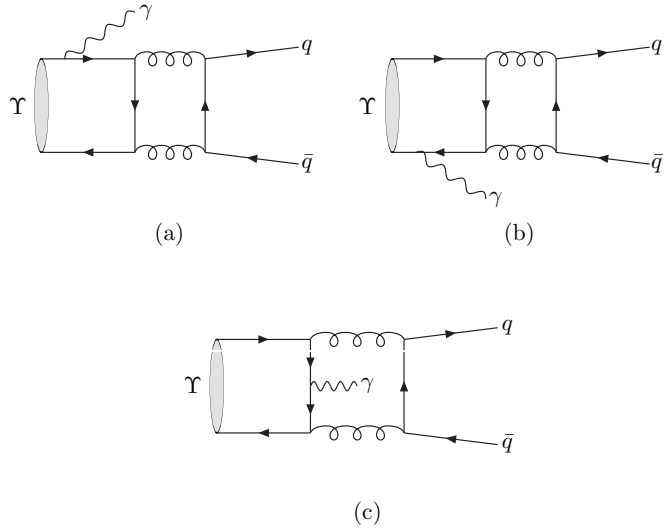}
\caption{Feynman diagrams for $\Upsilon\to \gamma + q +\bar q$ with
the color singlet $b\bar b$} \label{fvs}
\end{figure*}

The $q$ and $\bar q$ dijet production through $\Upsilon$ radiative
decay  at leading order is described by the process
 \bqa
 \Upsilon(P) \rightarrow \gamma g^{\ast} g^{\ast} \rightarrow \gamma(k_1)
 q(k_2) \bar{q}(k_3)
 \eqa
as shown in Fig.1 (another three diagrams with reverse light quark
lines are also taken into account).

In our work, all outgoing particles are taken to be on-mass-shell,
$k_1^2=k_2^2=k_3^2=0$ and $P^2=M_{\Upsilon}^2$. Here the light
quark $q$ (with momentum $k_2$) and antiquark $\bar q$ (with
momentum $k_3$) are taken to be massless, as for the photon (with
momentum $k_1$). We can see explicitly that there exist both
collinear and soft singularities in the loop integrals within the
region where the light quark and antiquark interact with two
gluons and are connected with a massless quark propagator. In
general, these soft and collinear mass singularities may be
separated by artificially introducing a quark mass and a gluon
mass. In the present case of dijet production, in order to define
two jets, the jet energy has to be higher than certain energy
scale, and the two jets have to be well separated by introducing
certain value for the angle between $q$ and $\bar q$. With these
two infrared cuts, the sensitivity of soft and collinear
singularities will become much weaker.



We use the NRQCD description for the $b\bar b$ bound state.  In
the static approximation, in which the relative motion of the
bottom quarks in the bound state is neglected, the decay amplitude
factorizes into a short distance amplitude $\mathcal{M}(b\bar{b}
\to \gamma q\bar{q})$, with $b \bar{b}$ in the color-singlet state
with zero relative momentum, and a long distance matrix element
which is related to $\mathcal{R}(0)$, the wave function of
$\Upsilon$ at the origin.

Within the static approximation, we write down the amplitude of
Fig.1(a) (for other related processes, see, e.g., Ref.\cite{berger})
 \bqa &&\mathcal{M}=\frac{C_Fg_s^4 eQ}{8 \sqrt{N_c \pi M}}
 \frac{\mathcal{R}(0)}{2 p\cdot k_1}\varepsilon^{(\ast)}_{\lambda,k_1}
 \int\frac{d^4 q}{(2\pi)^4}\bar{u}_i\gamma_{\mu} \pslash q\gamma_{\nu}v_j\\
 && \frac{Tr\big[\Pi_{1S_z}(P,M)\gamma^{\nu}(\pslash q+\pslash k_3-\pslash p+m)
 \gamma^{\mu}(\pslash p-\pslash k_1+m)\gamma^{\lambda}\big]}{q^2\ (q-k_2)^2\ (q+k_3)^2\ ((q+k_3-p)^2-m^2)}\nonumber. \eqa
where $p=P/2$, $m=M/2$,  $i$ and $j$ are color indexes, $g_s$ and
$e$ are the strong and electromagnetic couplings respectively, and
$Q$ is the magnitude of the bottom quark charge in units of $e$.
Charge conjugation invariance implies that the Feynman graphs are
symmetric under reversion of the fermion flow. All six amplitudes
contributing to the leading-order cross section are proportional
to the same color factor $1/(2\sqrt{N_c})\;\delta_{ab}$, and the
$\delta_{ab}$ combined with the color matrix of the right hand
fermion flow contribute a factor $C_F$ together with a color unit
matrix. Gauge invariance ensures that we may sum over the photon
spin by employing the substitutions
$\sum\varepsilon_\mu\varepsilon^{\ast}_\nu = - g_{\mu\nu}$. To sum
over the $\Upsilon$ spin, we use $ \sum \epsilon_{\Upsilon}^{\rho}
\epsilon_{\Upsilon}^{(\ast)\sigma} = - g^{\rho\sigma} +
P^{\rho}P^{\sigma}/M^{2}$.

In our calculation, we choose two parameters to describe the jets:
one is the minimum energy cut of the jet, which is, say, about
1~GeV, and the other is the angle between the two jet axes, which
is, say, about 36$^\circ$ (these values are just for
illustration). We choose the following region:
 \bqa
x\geq s_0\, ,\,\,\, |z|\leq z_0, \eqa

where x is the energy fraction of the jet ($k^0=m x$), and
$z$=$\cos\theta$, $\theta$ is angle between two jet axes.

Since we have taken some steps to separate the soft and collinear
divergences, we can complete the phase space integral in
4-dimension. The standard form of three body phase space in
4-dimension is:
 \bqa
d\Phi_3=\prod_{i=1}^3\frac{d^3 \vec{k}_i}{2 k_i^0(2 \pi)^3}(2
\pi)^4\delta^{4}(P-\sum_i k_i).
 \eqa
With the parameters
 \bqa
x_i=\frac{2 P k_i}{M^2}, \ \ \ \sum x_i=2,
 \eqa
we can get
 \bqa
d\Phi_3=\frac{M^2}{2(4 \pi)^3}\prod_{i=1}^{3}dx_i\delta(2-\sum x_i).
 \eqa
With the following constrains:
 \bqa
s_0\leq x_2\leq 1, \ s_0\leq x_3\leq 1,\ |\cos{\theta}|\leq z_0,
 \eqa
we can get the integration region of phase space: $-z_0$ $\leq$
$\cos{\theta}$ $\leq$ $z_0$ for $s_0$ $\leq$ $x_2$ $\leq$ $(2-2
s_0+r)/(2+s_0 z_0-s_0)$;~  $-z_0$ $\leq$ $\cos{\theta}$ $\leq$
$(2-2 s_0+r)/(s_0 x_1)+1-2/s_0$ for $(2-2 s_0+r)/(2+s_0 z_0-s_0)$
$\leq$ $x_2$ $\leq$ $(2-2 s_0+r)/(2-s_0 z_0-s_0)$, where
$r=$$m_q^2/m^2$. Here for numerical estimate we choose $s_0=0.2$
and $z_0=0.8$, corresponding to the jet energy cut 0.95~GeV and
the angle 36$^\circ$ between $q$ and $\bar q$.

In our calculation, we define $\langle
 \mathcal{O}_1^{\Upsilon(1S)}(^3S_1)\rangle$ $\simeq$ $\frac{9}{2 N_c}$
$|\mathcal{R}(0)|^2$\cite{braaten}, and take $m\approx
M/2=4.73$~GeV, and
 \bqa
\alpha_s=0.184,\,\,\,\,|\mathcal{R}(0)|^2
=7.12\,\,\,\,\mbox{GeV$^3$}\,, \eqa which are extracted from the
observed decay widths of $\Upsilon\to 3g$ and $\Upsilon\to e^+e^-$
with theoretical expressions for these decay widths including next
to leading order QCD radiative corrections. We set the gluon mass
as $10^{-6}$-- $10^{-8}$~GeV, together with the light quark mass
$m_q$ as $10^{-2}$-- $10^{-3}$~GeV, and find the numerical results
are stable as these masses tend to zero. Finally we find the
branching ratio for the dijet production in the color-singlet
sector to be \bqa Br(\Upsilon\to \gamma q\bar{q})=8.2\times
10^{-4}.\eqa

\section{Color octet contributions }

Next we take the color octet contributions into consideration.
With the velocity expansion in NRQCD the wave function of
$\Upsilon$ contains high Fock states with color-octet $b\bar b$
pair and associated soft gluons . We can see that the color-octet
$b\bar b$ operators $\mathcal{O}_8(^{3}S_1)$,
$\mathcal{O}_8(^{1}S_0)$, and $\mathcal{O}_8(^{3}P_J)$ will
contribute to the process $\Upsilon\to \gamma q\bar{q}$. The
corresponding diagrams are shown in Fig.2.

 \begin{figure*}[t]
\includegraphics[width=11cm]{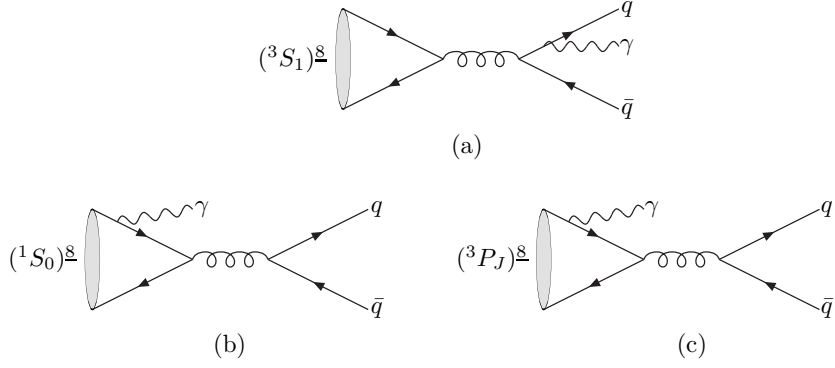}
\caption{Feynman diagrams with the color octet $b\bar b$ .}
\label{fvs}
\end{figure*}
The color octet operators e.g. $\mathcal{O}_8(^{3}S_1)$ and
$\mathcal{O}_8(^{1}S_0)$ are  given by
 \bqa
\mathcal{O}_8(^{3}S_1)&=& \psi^\dagger \mbox{\boldmath $\sigma$} T^a
\chi \cdot \chi^\dagger \mbox{\boldmath $\sigma$} T^a \psi
,\nonumber\\
\mathcal{O}_8(^{1}S_0)&=& \psi^\dagger T^a \chi \, \chi^\dagger T^a
\psi.
 \eqa
Differing from the color singlet case, the color projection of the
octet is
 \bqa
\langle3 i;\overline{3} j|\ 8 a\rangle\ =\ \sqrt{2}\ T^a_{ij}.
 \eqa
The spin projection operator e.g. $\Pi_{00}(P,M)$ is \cite{Kuhn}:

 \bqa \Pi_{00}(P,M)\ =\ -\frac{1}{\sqrt{4\ M}}(\pslash P+M)\,\gamma_5. \eqa
We can write down the decay amplitudes of the color octet $b\bar
b$ as shown in Fig.2(a),(b), and (c) respectively:

\begin{widetext}
 \bqa
 \mathcal{M}_a &=& -i \sqrt{2}\ g_s^2 e\ Q_q\ Tr[T^a\ T^b]\varepsilon^{(\ast)}_{\lambda}(k_1)
 \frac{Tr[\Pi_{1S_z}(P,M) \gamma^{\mu}]}{4 m^2}\ \overline{q}\gamma^{\lambda}
 \frac{2\pslash{p}-\pslash{k_3}}{(2 p-k_3)^2}T^a\gamma_{\mu}q,\nonumber\\
 \mathcal{M}_b &=& i \sqrt{2}\ g_s^2 e\ Q_b\ Tr[T^a\ T^b]\varepsilon^{(\ast)}_{\lambda}(k_1)
 \frac{Tr[\Pi_{00}(P,M) \gamma^{\mu}(\pslash{p}-\pslash{k_1})\gamma^{\lambda}]}{(k_2+k_3)^2[(p-k_1)^2-m^2]}\
 \overline{q}T^a\gamma_{\mu}q,\nonumber\\
 \mathcal{M}_c &=& -i \sqrt{2}\ g_s^2 e\ Q_b\ Tr[T^a\ T^b]\varepsilon^{(\ast)}_{\lambda}(k_1)
 \frac{\overline{q}T^a\gamma_{\mu}q}{(k_2+k_3)^2[(p-k_1)^2-m^2]}\bigl[Tr[\Pi^{\alpha}_{1S_z}(P,M) \gamma^{\mu}(\pslash{p}-\pslash{k_1})\gamma^{\lambda}]
 \nonumber\\
 &+&Tr[\Pi_{1S_z}(P,M) \gamma^{\mu}\gamma^{\alpha}\gamma^{\lambda}]
 +\frac{2\ k_1^{\alpha}}{(p-k_1)^2-m^2}
 Tr[\Pi_{1S_z}(P,M) \gamma^{\mu}(\pslash{p}-\pslash{k_1})\gamma^{\lambda}]\bigr], \eqa
\end{widetext}
where  $Q_q$ is the light quark charge. Then we can express the
decay width as functions of the jet energy cut $s_0$ and the jet
angle cut $z_0$. With $s_0$=0.2 (corresponding to the jet energy
cut 0.95~GeV) fixed,  the decay width (in arbitrary units) as
functions of $z_0$ are shown in Figs.4-6. We see that the decay
width increases nearly linearly as $z_0$ increases when $z_0<0.8$,
whereas when $z_0>0.8$ the decay width will increase rapidly. This
may imply that taking the jet angle cut to be $z_0<0.8 $~
($\theta>36^\circ$) is a reasonable choice to avoid the
sensitivity of infrared singularities.

Choosing $\alpha_s(M_{\Upsilon})$=$0.184$ and $\alpha$=$1/137$,
and including color-singlet and various color-octet contributions,
we can write down the branching ratio as
 \bqa
 Br(\Upsilon \to \gamma q\bar{q})&=&2.4\times 10^{-4}\langle
 \Upsilon|\mathcal{O}_1(^3S_1)|\Upsilon\rangle\nonumber\\
 &+&0.061\times\langle \Upsilon|\mathcal{O}_8(^3S_1)|\Upsilon\rangle\nonumber\\
  &+&0.084\times\langle \Upsilon|\mathcal{O}_8(^1S_0)|\Upsilon\rangle\nonumber\\
 &+&0.043\times\langle
 \Upsilon|\mathcal{O}_8(^3P_0)|\Upsilon\rangle
 \eqa
 \begin{figure}[t]
\includegraphics[width=6.5cm]{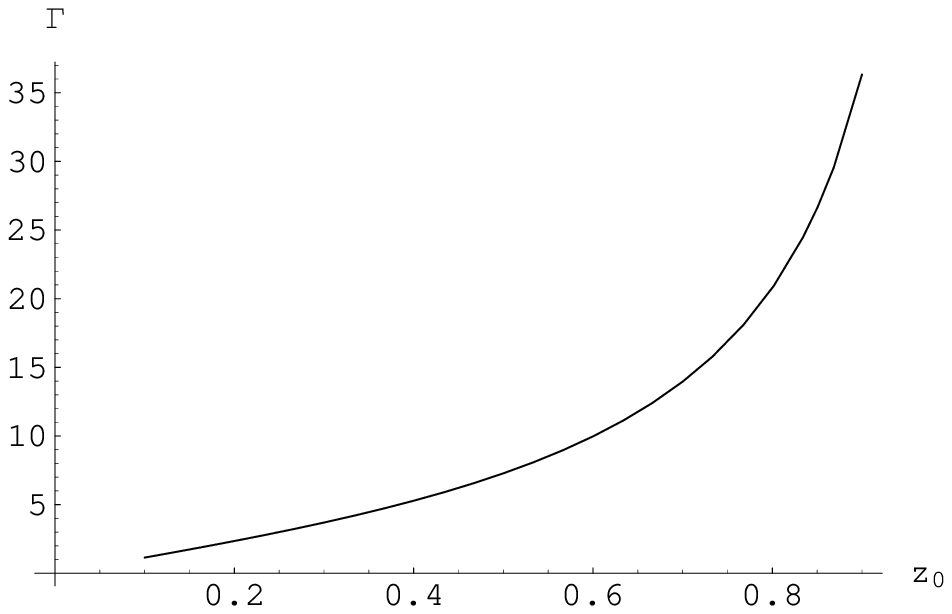}
\caption{Decay width (in arbitrary units) as a function of the angle
cut from the color octet $^{3}S_{1}$ operator.} \label{fvs}
\end{figure}

 \begin{figure}[t]
\includegraphics[width=6.5cm]{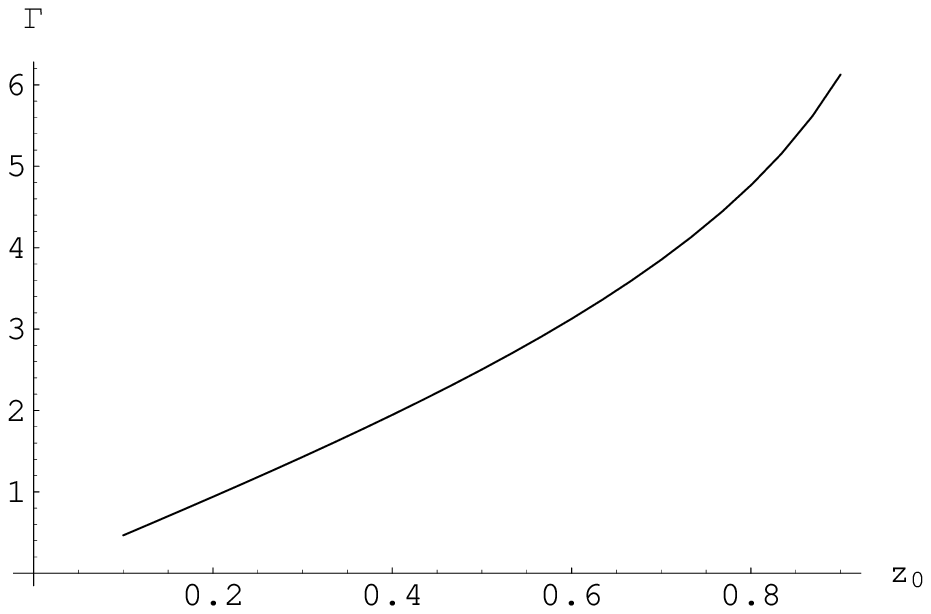}
\caption{Decay width (in arbitrary units) as a function of the angle
cut from the color octet $^{1}S_{0}$ operator.} \label{fvs}
\end{figure}

 \begin{figure}[t]
\includegraphics[width=6.5cm]{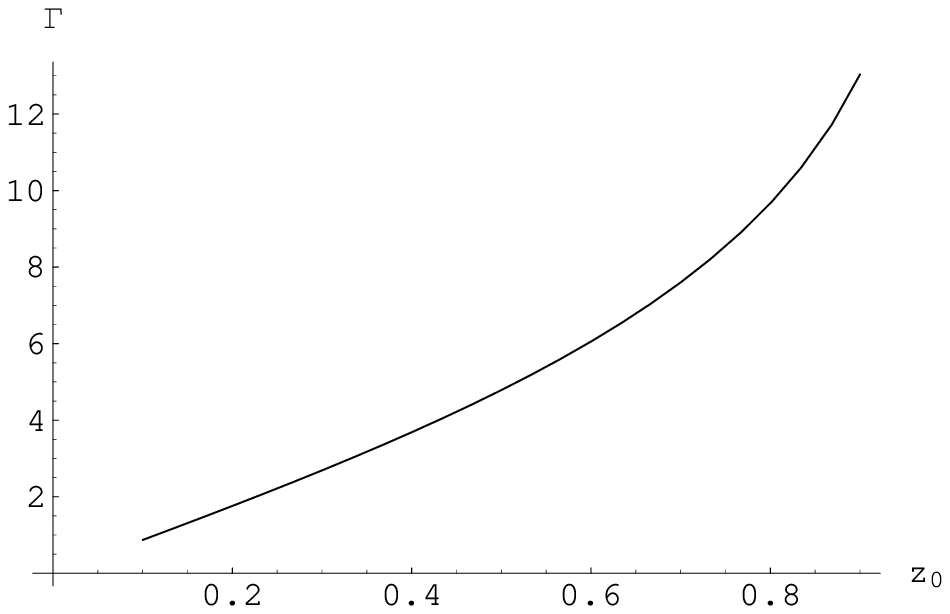}
\caption{Decay width (in arbitrary units) as a function of the angle
cut from the color octet $^{3}P_{J}$ operator.} \label{fvs}
\end{figure}


For numerical results of the decay branching ratio, the main
uncertainty comes from the estimates of the color-octet matrix
elements. The most direct way is to use the velocity scaling rules.
However, it is shown in Refs.\cite{rgeq, maltoni2} that by using
renormalization group equations the obtained color octet matrix
elements are smaller. Nonetheless, in our calculation we still
choose velocity scaling rules with $v^2=0.08$ for
$\Upsilon$\cite{fleming3} as a roughly estimate. Another estimate
can be made from the $\Upsilon$ production at the Fermilab Tevatron
using NRQCD \cite{braaten}, and the extracted matrix elements are
shown in Table.I (for more discussions on the color octet matrix
elements, see e.g. Refs.\cite{co1,co3}).

As the first estimate, we set the color octet $^1S_0$ and $^3P_0$
matrix elements to be zero, and use the following velocity scaling
rule relation for the $^3S_1$ matrix element\cite{fleming3}:
 \bqa
 \langle \Upsilon|\mathcal{O}_8(^3S_1)|\Upsilon\rangle&=&v^4\langle \Upsilon|\mathcal{O}_1(^3S_1)|\Upsilon\rangle.
 \eqa
and we then get the color octet  $^3S_1$ contribution to be
 \bqa
Br(\Upsilon\to \gamma q\bar{q})&=&1.33\times 10^{-3},\eqa which is
larger than the color singlet contribution by a factor of 1.6.
\begin{table}[tb]
\begin {center}
\begin{tabular}{|c|c|c|}
 \hline
$\langle \mathcal{O}_8^{\Upsilon(1S)}(^3S_1)\rangle$&$2.0\pm4.1^{-0.6}_{+0.5}$&$3.0\pm3.8^{+0.2}_{-0.1}$\\
$\langle \mathcal{O}_8^{\Upsilon(1S)}(^1S_0)\rangle$&$13.6\pm6.8^{+10.8}_{-7.5}$&$0$\\
$\frac{5}{m_b^2}\langle \mathcal{O}_8^{\Upsilon(1S)}(^3P_0)\rangle$&$0$&$13.9\pm7.1^{+11.4}_{-8.0}$\\
\hline
\end{tabular}
\caption{Color octet matrix elements for $\Upsilon$ (in units of
$10^{-2}$ $\mbox{GeV}^3$), taken from Table V in
Ref.\cite{braaten}.}
 \label{table1}
\end {center}
\vspace{-0.5cm}
\end{table}

Next, we use the results from the analysis of $\Upsilon$
production at the Fermilab Tevatron, and we adopt the assumption
that the production matrix elements are 3 times larger than the
annihilation matrix elements just like the case of color singlet
matrix elements. Note that the color octet matrix element $\langle
\mathcal{O}_8^{\Upsilon(1S)}(^1S_0)\rangle$ and $\frac{5}{m^2}\
\langle\mathcal{O}_8^{\Upsilon(1S)}(^3P_0)\rangle$ can not be
determined independently, and the values shown in Table.I from
\cite{braaten} are a sort of simplification. We use the central
values in the first column (the $^3P_0$ matrix element is zero),
and in the second column (the $^1S_0$ matrix element is zero)
respectively to calculate the decay branching ratio, and the
results are:
 \bqa
Br(\Upsilon\to \gamma q\bar{q})=4.21\times 10^{-3},\nonumber\\
Br(\Upsilon\to \gamma q\bar{q})=9.31\times 10^{-3}.\eqa
The first branching ratio is contributed from
$\mathcal{O}_8(^3S_1)$ and $\mathcal{O}_8(^1S_0)$, and the second
is from $\mathcal{O}_8(^3S_1)$ and $\mathcal{O}_8(^3P_J)$. These
branching ratios are really big!  To compare with the contribution
from color singlet, we draw two group curves based on the latter
calculation. Fig.6 shows angular distributions of the decay
branching ratio from different channel contributions. We see that
both in the large angle region and the small angle region, color
octet gives  greater contribution than singlet does. If
experimentally the jet angular distribution can be measured  then
we might find useful hints of color octet contributions.

Fig.7 shows the branching ratio angular distributions for each
individual color octet operators. Here we choose $\langle
\mathcal{O}_8^{\Upsilon(1S)}(^3S_1)\rangle$=$2.0\times
10^{-2}~GeV^3$. The values for other two operators are from
Table.I. These curves are also useful in distinguishing between
contributions of different color octet operators.

 \begin{figure}[t]
\includegraphics[height=6.5cm]{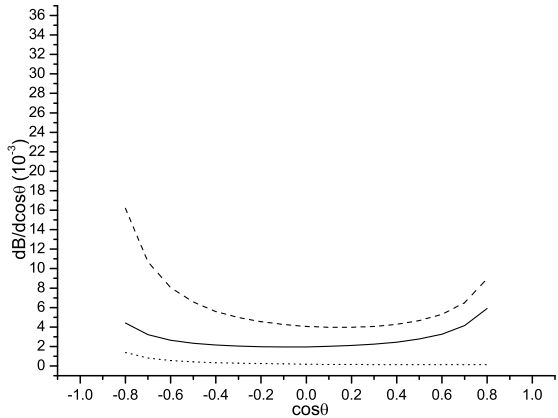}
\caption{Decay branching ratio angular distributions. The dotted
line is for the color singlet contribution, dashed line for
$^{3}S_{1}$ and $^{3}P_{J}$ color octet contributions, and solid
line for $^{3}S_{1}$ and\ $^{1}S_{0}$ color octet contributions.}
\label{fvs}
\end{figure}

 \begin{figure}[t]
\includegraphics[height=6.5cm]{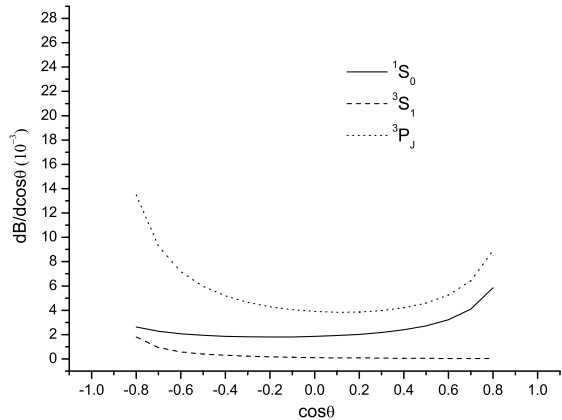}
\caption{Angular distributions for different color octet
operators.} \label{fvs}
\end{figure}



\section{Summary}

In this paper, we studied two mechanisms i.e. the color-singlet
and color-octet mechanisms that contribute to the process
$\Upsilon \to \gamma q\bar{q}$. With the help of FeynCalc
\cite{FeynCalc} and LoopTools\cite{LoopTools}, we calculated
numerically the radiative decay rates by introducing the energy
cut for the $q$ and $\bar q$ jets, and the angle cut between $q$
and $\bar q$ jets. These cuts correspond to a minimum jet energy
of about $1 \mbox{GeV}$ and the angle between two jets larger than
$\theta=36.8^{\circ}$. Experimentally, these cuts are necessary
for the two jets measurement. Theoretically, introducing these
cuts greatly reduces the sensitivity to the soft and collinear
singularities in the loop integrals.  Furthermore, from our
studies we find that change of energy cut affects the result very
mildly, and change of the angle cut brings a nearly linear effects
in the region $|cos\theta|<0.8$.  This implies that in this region
those large logarithms which are introduced by regularizing the
soft and collinear divergences only give small contributions and
may then be ignored. Therefore, our numerical results should be
reasonable.

Based on our calculation in NRQCD, we predict the branching ratio of
$\Upsilon \to \gamma q\bar{q}$ to be $8.2\times 10^{-4}$ if
contributions from the color singlet $b\bar b$ are considered only .

On the other hand, the color octet $b\bar b$ operators may play an
important role in this process. When we take them into
consideration, the branching ratio could be enhanced by about an
order of magnitude [see Eq.(18)]. This result is very encouraging.
In this sense, the process $\Upsilon \to \gamma q\bar{q}$ may
provide an interesting test for the color octet mechanism in NRQCD.
However, large uncertainties in the estimate of color-octet matrix
elements would prevent us from making definite predictions for this
process. A better understanding of the color octet matrix elements
in the future will certainly improve our predictions.

We hope that the measurement on $\Upsilon \to \gamma q\bar{q}$
will clarify the theoretical issues presented in this paper.

\begin{acknowledgments}
We thank E. Berger for bringing their earlier work on inelastic
photoproduction of $\Upsilon$ (Ref.\cite{berger}) to our attention.
We would also like to thank Ce Meng for valuable discussions. This
work was supported in part by the National Natural Science
Foundation of China (No 10421503), and the Key Grant Project of
Chinese Ministry of Education (No 305001).

\end{acknowledgments}


\end{document}